\documentclass[onecolumn, 11pt]{article}
\usepackage[margin=1in]{geometry}
\usepackage{setspace}
\usepackage{amsmath}
\usepackage{subcaption}

\usepackage{graphicx}
\usepackage{epstopdf}
\usepackage{amssymb}

\usepackage[T1]{fontenc}
\usepackage[utf8]{inputenc}
\usepackage{authblk}

\usepackage[numbers]{natbib}
\usepackage[OT4]{fontenc}
\usepackage[utf8]{inputenc}

\newcommand{\Exp}{{\rm I\hspace{-0.8mm}E}}

\newcommand{\Var}{{\bf Var}}

\newcommand{\Corr}{{\bf Corr}}

\newcommand{\iz}{{\rm \rlap Z\kern 2.2pt Z}}

\newcommand{\proof}{\noindent {\bf Proof:} \ }

\newtheorem{theorem}{Theorem}

\newtheorem{proposition}{Proposition}

\author[1]{Zbigniew Michna\footnote{zbigniew.michna@ue.wroc.pl\\ Tel/fax: +48713680335}}
\author[2]{Stephen M. Disney\footnote{DisneySM@cardiff.ac.uk\\ Tel: +44(0)2920876310}}
\author[3]{Peter Nielsen\footnote{peter@m-tech.aau.dk\\ Tel: +4599408932}}
\affil[1]{Department of Mathematics and Cybernetics, Wroc{\l}aw University of Economics, Poland.}
\affil[2]{Logistics Systems Dynamics Group, Cardiff Business School, Cardiff University, Wales.}
\affil[3]{Department of Mechanical and Manufacturing Engineering, Aalborg University, Denmark.}

\title{\bf\LARGE {The impact of stochastic lead times \\ on the bullwhip effect under correlated demand \\
and moving average forecasts}}

\date{\today}

\begin{document}
\onehalfspacing
\maketitle

\bibliographystyle{abbrv}

\begin{abstract}
We quantify the bullwhip effect (which measures how the variance of replenishment orders is amplified as the orders move up the supply chain)  when random demands and random lead times are estimated using the industrially popular moving average forecasting method. We assume that the lead times constitute a sequence of independent identically distributed random variables and correlated demands are described by a first-order autoregressive process.

We obtain an expression that reveals the impact of demand and lead time forecasting on the bullwhip effect. We draw a number of conclusions on the bullwhip behaviour with respect to the demand auto-correlation and the number of past lead times and demands used in the forecasts. Furthermore, we find the maxima and minima in the bullwhip measure as a function of the demand auto-correlation.

\vspace{5mm}
{\it Keywords: supply chain, bullwhip effect,
order-up-to replenishment policy, AR(1) demand, stochastic lead time, moving average forecasting method.}
\end{abstract}

\section{Introduction}
The variability of replenishment orders often increases as they flow upstream in supply chains. This phenomenon is known as the bullwhip effect and has been discussed in the economics and operations management literature for 100 and 50 years, respectively -- see Mitchell \cite{mi:23} and Forrester \cite{fo:58}. The celebrated works of Lee et al., \cite{le:pa:wh:97a} and \cite{le:pa:wh:97b} promoted this problem to the forefront of the supply chain and operations management field. Wang and Disney \cite{wa:di:15} provide a recent literature review of the bullwhip field, categorising contributions according to the five causes of bullwhip of Lee et al.: demand forecasting, non-zero lead time, supply shortage, order batching and price fluctuation. Of particular importance to this paper are the results of Chen et al., \cite{ch:dr:ry:si:00a}, \cite{ch:dr:ry:si:00b} and Dejonckheere et al., \cite{de:di:la:to:03}.  These contributions investigate the bullwhip consequences of using the moving average forecasting method inside the order-up-to (OUT) replenishment policy.

Recently Michna and Nielsen \cite{mi:ni:13} identified another critical cause of the bullwhip -- the forecasting of lead times. While the issue of stochastic lead times in bullwhip studies has not been intensively investigated, Michna and Nielsen \cite{mi:ni:13} and Michna et al., \cite{mi:ni:ni:14} provide a recent literature review of this problem. Of particular importance is the work of Duc et al., \cite{du:lu:ki:08} and Kim et al., \cite{ki:ch:ha:ha:06} where the impact of stochastic lead times on bullwhip is quantified.  These works characterise the impact of random lead times on the bullwhip effect via mean values and variances. However, they do not consider the consequences of having to estimate the lead time distribution (a.k.a. lead time forecasting). As identified by Michna and Nielsen \cite{mi:ni:13} and Michna et al., \cite{mi:ni:ni:14} this can be a significant cause of the bullwhip effect. In Duc et al., \cite{du:lu:ki:08} lead times are assumed to be stochastic and drawn from a known distribution and thus are not forecasted when placing an order. Kim et al., \cite{ki:ch:ha:ha:06} used the moving average technique to forecast lead time demand, as did Michna et al., \cite{mi:ni:ni:13}.

The influence of stochastic lead time on inventory is a more established field and we refer to the work of Bagchi et al., \cite{ba:ha:ch:86}, Chaharsooghi et al., \cite{ch:he:10}, Song \cite{so:94a} and \cite{so:94b}, and Zipkin \cite{zi:86}. Stochastic lead time inventory research can be classified into two general streams: those with order crossovers and those without crossovers. An order crossover happens when replenishments are received in a different sequence from which the orders were placed (see e.g. Bischak et al., \cite{bi:ro:si:bl:14}, Bradley and Robinson \cite{br:ro:05}, Disney et al., \cite{di:ma:wa:wa:16} and Wang and Disney \cite{wa:di:15}). Disney et al., \cite{di:ma:wa:wa:16} consider the safety stock and inventory cost consequences of using the OUT and proportional order-up-to (POUT) replenishment policies under i.i.d. demand.  They show that the POUT policy is always more economical than the OUT policy when order-crossover is present. Wang and Disney \cite{wa:di:15} show that the POUT policy outperforms the OUT policy in the presence of order crossovers in the sense of minimizing inventory variance when demand is an Auto-Regressive, Moving Average process with $p$ auto-regressive terms and $q$ moving average terms, ARMA(p,q).

The papers of Boute at el. \cite{bo:di:la:ho:08}, \cite{bo:di:la:ho:09},  \cite{bo:di:la:ho:14} investigate endogenous lead times in supply chains. Endogenous lead times are dependent on the state of the system as they are function of the previous orders. Here the supplier is modelled as a queue and orders are processed on a first come, first served basis, hence there is no order-crossover. However, as the sojourn time in the queue increases in the variance of the demand placed on the manufacturer, a lead time reduction can be obtained by smoothing the replenishment orders. This lead time reduction can potentially reduce safety stock requirements. Hum and Parlar \cite{hu:pa:14} also model lead times using queueing theory, analyzing the proportion of demand that can be met within a specific lead time.

\begin{figure}[!h]
\begin{center}
\includegraphics[width=11cm]{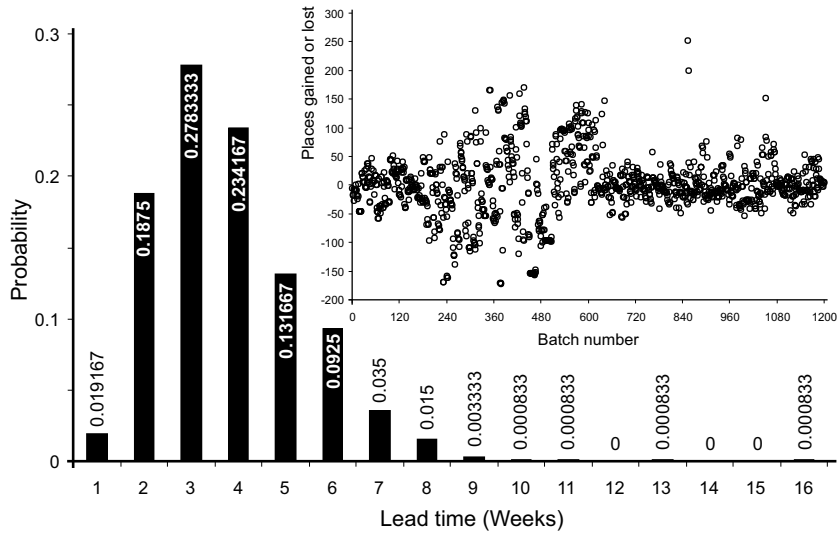}
\caption{Stochastic lead times and order-crossovers observed in a measuring equipment supplier}\label{Fig1}
\end{center}
\end{figure}

\begin{figure}[!h]
\begin{center}
\includegraphics[width=11cm]{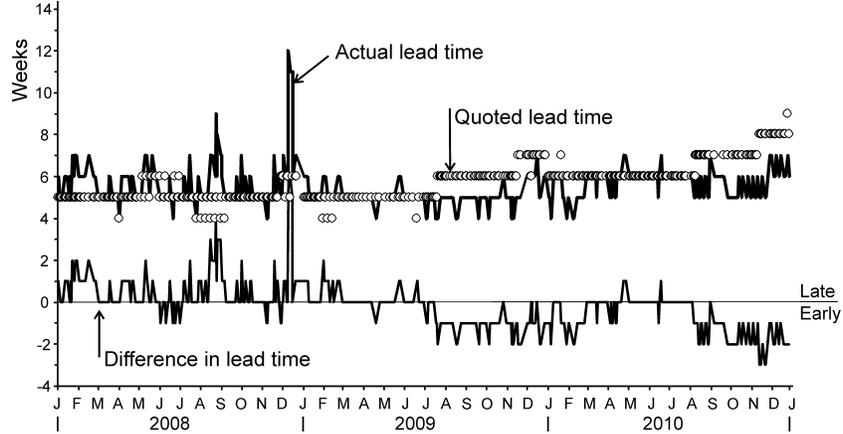}
\caption{The quoted and actual lead times in a global shipping lane}\label{Fig2}
\end{center}
\end{figure}

We have observed that stochastic lead times and order-crossovers are quite common within factories (see Fig. \ref{Fig1}). The data represents a single, high volume, product from a supplier of industrial measuring and testing equipment. The distribution of the lead times is discrete and aggregated into weekly buckets to reflect the actual practice of creating weekly production plans using the OUT policy (for more information on why this is so, we refer to the assumptions and modelling choices discussed later in this section). Fig. \ref{Fig1} also highlights the number of queue positions each production batch gained or lost between the two lists of date sorted production releases and production completions. As this manufacturer manually moved totes of products between process steps within its job shop, a large number of order-crossovers is present. Disney et al., \cite{di:ma:wa:wa:16} present similar findings in global supply chains (see Figs. \ref{Fig1} and \ref{Fig2} of \cite{di:ma:wa:wa:16}), where stochastic lead times and order crossovers could be observed in global shipping lanes.  Here containers could also gain or lose positions in the date ordered list of dispatches and receipts. We also observe differences in quoted (at the time of shipping) and actual (realised when the container arrives) lead times in global shipping lanes (see Fig. \ref{Fig2}).

We consider a model where a supply chain member (who could be a retailer, manufacturer, or supplier for example, but we call a manufacturer for convenience) observes both random demands from his customer and random lead times from his supplier which we assume to be exogenous (that is, they are independent of all other system states). The manufacturer generates replenishment orders to maintain inventory levels by projecting his customers' future demands over his supplier's lead time, accounting for both the available inventory and the open orders in the replenishment pipeline.

This research differs from previous research in several ways. Most importantly we show that lead time forecasting is a major cause of bullwhip when demands are auto-correlated. This confirms and extends the results of Michna and Nielsen \cite{mi:ni:13}. We also quantify the impact of the stochastic auto-correlated demands and stochastic lead times on the bullwhip effect under the assumption that demands and lead times are forecasted separately using moving averages. Furthermore, we investigate the bullwhip effect as a function of the demand auto-correlation, the characteristics of the lead time distribution and the number of past demands and the delay parameter in the moving average lead time forecasts. The bullwhip conclusions differ depending on how the parameters are combined. We find maxima and minima in the bullwhip metric as a function of the demand auto-correlation.

Moreover our main result contains, as special cases, the bullwhip formulas of Chen et al., \cite{ch:dr:ry:si:00a} (a constant lead time) and Th. 1 in Michna and Nielsen \cite{mi:ni:13} (mutually independent demands). The formulation presented in this research involves more parameters, is more general, and allows us to understand more intricate supply chain settings.

Our major assumptions and modelling choices are summarised as follows:
\begin{enumerate}
\item[a)] The supply chain consists of two stages -- a manufacturer who receives client's demands and deliveries from a supplier (or manufacturing process).
\item[b)] A periodic replenishment system exists where the demands, $D_t$, are satisfied and previous orders placed are received during a time period, indexed by the subscript $t$. At the end of the period, the inventory level, demand and lead times of received orders are observed and a new replenishment order, $q_t$, is placed. The length of the period could be an hour, day, week or month, but in our experience it is often a week in manufacturing contexts. Note that the receipt of an order is observed only at the end of the period and the lead time is a non-negative integer. An order with zero lead time would be received instantaneously after the order was placed, but its receipt would only be incorporated into the order made at the end of the next period due to the sequence of events delay.
\item[c)] The demand constitutes an autoregressive model of order one, AR(1).  We have elected to use the AR(1) model as it is the simplest demand process with autocorrelation, a feature commonly observed in real demand patterns, Lee et al., \cite{le:so:ta:00}. It is also a frequently adopted assumption in the bullwhip literature (e.g. in Chen at el. \cite{ch:dr:ry:si:00a} and \cite{ch:dr:ry:si:00b}, Duc et al., \cite{du:lu:ki:08} and Lee et al., \cite{le:so:ta:00}), allowing comparison of our new results to established theory.
\item[d)] The lead times $L_t \in \mathbb{N}_0$ constitute a sequence of independent identically distributed (iid) random variables which are independent of all system states, including the manufacturer's demand. Moreover we assume that lead times are bounded  (e.g.  $L_t \leq L^+$ periods) and that the lead time forecasts are based on lead time information that is at least $L^+$ periods old. This allows use to create lead time forecasts that are unbiased.  For example, if we based our lead time forecasts on the most recent lead time information (which we observe when we receive orders), some of the orders placed would still be open (not yet received) and our lead time estimates would only be based on those orders with short lead times.  Basing our lead time estimates on data that is at least $L^+$ periods old is possible as lead times are assumed to be temporally independent and thus constitute a valid dataset for forecasting all future lead times.  Practically this approach has the desirable characteristic that we can base our lead time estimates on realised lead times, rather than quoted lead times from the supplier or shipper, see Fig. \ref{Fig2}.  Furthermore, for ease of data organisation (and modelling) we can retrospectively assign the lead time of an order to the period the order was generated in our database (simulation).
\item[e)] The OUT policy is used to generate the orders placed onto the supplier. The OUT policy is industrially popular as it is commonly available native in many ERP/MRP systems.  It has also been studied extensively in the academic literature (see e.g. Bishak at el. \cite{bi:ro:si:bl:14}, Chen at el. \cite{ch:dr:ry:si:00a} and \cite{ch:dr:ry:si:00b}, Dejonckheere at el. \cite{de:di:la:to:03} and \cite{de:di:la:to:04}, Duc at el. \cite{du:lu:ki:08} and Kim at el. \cite{ki:ch:ha:ha:06}).  The OUT policy is also the optimal linear replenishment policy for minimizing inventory holding and backlog costs if orders do not cross (see Kaplan \cite{ka:70} and Wang and Disney \cite{wa:di:15}).
\item[f)] The manufacturer predicts the future demands over future lead times based on predictions generated using the moving average forecasts of past demand and observations of the lead times of previously received orders. Thus, the forecast of lead time demand is as follows
\begin{equation}\label{ltdf}
\widehat{D_t^L}=\sum_{i=0}^{\widehat {L_t}-1}\widehat{D}_{t+i}\,,
\end{equation}
where $\widehat{L_t}$ is the forecast of the lead time of the next order made at the beginning of period $t$ and
$\widehat{D}_{t+i}$ denotes the forecast for a demand for the period $t+i$ made at the beginning of a period $t$.
\end{enumerate}

As Michna and Nielsen \cite{mi:ni:13}, the novel aspect of our approach is the last point f) and differs from much of the previous literature. For example, Duc et al., \cite{du:lu:ki:08} assume the lead time of the order placed at time $t$ is known when placing order leading to
$$
\widehat{D_t^L}=\sum_{i=0}^{L_t-1}\widehat{D}_{t+i}\,,
$$
However, we assume the manufacturer would not know the value of $L_t$ until that order has been completed (arrived, received).

In Kim et al., \cite{ki:ch:ha:ha:06} the lead time demand is predicted with
$$
\widehat{D_t^L}=\frac{1}{n}\sum_{i=1}^{n}D_{t-i}^L\,,
$$
where $D_{t-i}^L$ is the past known (realized) lead time demand.

A different approach was taken by Bradley and Robinson \cite{br:ro:05} and Disney et al., \cite{di:ma:wa:wa:16} where it is assumed beforehand that the lead time distribution is known.  That is, the lead time distribution can be observed from previous realisations of the lead time.

In our approach we show that the bullwhip effect measure contains new components depending on the lead time forecasting parameter, and the correlation coefficient between demands.  This was not quantified in Michna and Nielsen \cite{mi:ni:13}, neither was it included in the study of ARMA(p,q) demand in Wang and Disney \cite{wa:di:15}. These new terms amplify the value of the bullwhip measure and are evidence that lead time estimation in itself is a significant cause of the bullwhip effect, perhaps equally as important as demand forecasting.

\section{Supply chain model}
We want to consider temporally dependent demands and the simplest way to achieve this is to model a manufacturer observing periodic customer demands, $D_t$, constituting of a stationary first-order autoregressive, AR(1), process,
\begin{equation}\label{ard}
D_t=\mu_D+\rho (D_{t-1}-\mu_D)+\epsilon_t\,,
\end{equation}
where $|\rho|<1$ and $\{\epsilon_t\}_{t=-\infty}^\infty$ is a sequence of independent
identically distributed random variables such that $\Exp (\epsilon_t)=0$ and $\Var(\epsilon_t)=\sigma^2$.
Under the stationarity assumption it can be easy found that $\Exp (D_t)=\mu_D$, $\Var (D_t)=\sigma^2_D=\frac{\sigma^2}{1-\rho^2}$ and $\Corr(D_t, D_{t-k})=\rho^k$ (see for example, Chen et al., \cite{ch:dr:ry:si:00a} and Duc et al., \cite{du:lu:ki:08}). The distribution of $D$ can be arbitrary but its second moment must be finite.

A random lead time $L_t$ is assigned to each order at the beginning of time $t$. It is observed and used to predict future lead time when the order is received. The random lead times $\{L_t\}_{t=-\infty}^\infty$ are mutually iid random variables that is also assumed in Duc et al., \cite{du:lu:ki:08}, Kim et al., \cite{ki:ch:ha:ha:06}, Robinson et al., \cite{ro:br:th:01} and Disney et al., \cite{di:ma:wa:wa:16}. The expected value of the discrete lead times is $\Exp L_t=\sum_{i=0}^{L^+}{i p_i}=\mu_L$ where $p_i$ is the probability that the lead time is $i$ periods long, $\Var L_t=\sum_{i=0}^{L^+}{p_i (i-\mu_L)^2}=\sigma_L^2$.
We do not impose any assumptions on the distribution of $L$ but that its second moment is finite and $L$ is non-negative.  The sequences $\{D_t\}_{t=-\infty}^\infty$ and $\{L_t\}_{t=-\infty}^\infty$ are mutually independent.

The lead time demand at the beginning of a period $t$ is defined as follows
\begin{equation}\label{ltd}
D_t^L=D_t+D_{t+1}+\dots...+D_{t+L_t-1}=
\sum_{i=0}^{L_t-1}D_{t+i}\,,
\end{equation}
which reflects the demand over the lead time.
At the beginning of a period $t$ the manufacturer does not know this value of $L_t$ so he must forecast its value before calculating his replenishment order (see (\ref{ltdf})).

Let us notice that there is a dependency between $\widehat{D_t^L}$ and $\widehat{L_t}$ due to (\ref{ltdf}).  That is, the lead time demand forecast is a function of past lead times. Employing the moving average forecast method with the delay parameter $n\geq 1$ for demand forecasting we get
\begin{equation}\label{df}
\widehat{D}_{t+j}=\frac{1}{n}\sum_{i=1}^n D_{t-i}\,,
\end{equation}
where $j=0,1,\ldots$ and $D_{t-i}$ $i=1,2,\ldots, n$ are previous demands which have been observed at the
beginning of period $t$. Here we use a simple moving average method. Thus the $j$-period ahead forecast of demand is a moving average of previous demands.  Note all future forecasts, regardless of $j$, are straight line predictions of the current forecast. Clearly this is not an optimal, minimum mean squared error, forecast for AR(1) demand. However, it does reflect common industrial practice as the moving average forecast is available in many commercial ERP systems and can be readily incorporated into spreadsheets by analysts. It has also been studied from a theoretical basis (see Chen et al., \cite{ch:dr:ry:si:00a}, Dejonckheere et al., \cite{de:di:la:to:03}, Kim and Ryan \cite{ki:ry:03}, Chatfield at el., \cite{ch:ki:ha:ha:04} and Chatfield and Hayya \cite{ch:ha:07}).

The manufacturer also predicts a lead time but here he has to be careful because the previous orders cannot be completely observed. Precisely, using the moving average forecast method with $m\geq 1$ for lead time forecasting we obtain
\begin{equation}\label{ltf}
\widehat{L_t}=\frac{1}{m}\sum_{i=1}^m L_{t-i-L^+}\,,
\end{equation}
where $L_{t-i-L^+}$ $i=1,2,\ldots, m$ are lead times which are guaranteed to have been observed by the manufacturer at the beginning of a period $t-i$ (or earlier) as they are at least $L^+$ periods old (see item d of our discussion of assumptions in $\S$1). Knowing the average lead time (in practice estimating it) we are able to find the average unrealized orders (see  Robinson et al., \cite{ro:br:th:01} and Disney et al., \cite{di:ma:wa:wa:16}).
However our procedure of collecting lead times avoids bias resulting from the open orders with long lead times that may not have been received when we make the lead time forecast.
Thus by  (\ref{ltdf}), (\ref{df}) and (\ref{ltf}) we propose the following forecast for a lead time demand (see also Michna and Nielsen \cite{mi:ni:13}).
\begin{equation}\label{eltd}
\widehat{D_t^L}=\widehat{L_t}\widehat{D_t}=
\frac{1}{mn}\sum_{i=1}^n D_{t-i}\sum_{i=1}^m L_{t-i-L^+}\,.
\end{equation}
It is easy to notice that (\ref{eltd}) is a slight modification of (\ref{ltdf}) when demands and lead times are predicted using the moving average method.
The motivation for the lead time demand forecasting given in  (\ref{eltd}) is also the fact
that $\Exp D_t^L=\Exp L\Exp D$ see  (\ref{ltd}) (under the assumption that demands and lead times are mutually independent) and employing the natural estimators of $\Exp L$ and $\Exp D$ we arrive at  (\ref{eltd}).
Eq. (\ref{eltd}) has previously been used by Chatfield et al., \cite{ch:ki:ha:ha:04}
in a simulation study that highlighted the relationship between lead time forecasting and the bullwhip effect.

We assume that the manufacturer uses the OUT policy. Let $S_t$ be the desired inventory position at the beginning of a period $t$,
\begin{equation}\label{st}
S_t=\widehat{D_t^L}+\mbox{\it TNS}\,,
\end{equation}
where $\mbox{\it TNS}$ is a constant, time invariant, target net stock (safety stock), set to achieve a desired level of availability or to minimize a set of unit inventory holding ($h$) and unit backlog ($b$) costs via the newsvendor principle, Silver et al., \cite{si:pi:pe:98}.  It is often assumed, in constant lead time scenarios, that the demand and the inventory levels, are normally distributed and thus
$$
\mbox{\it TNS}=z\widehat{\sigma_t} \text{, } z=\Phi^{-1}\left(\frac{b}{b+h}\right)
$$
holds, where $\Phi^{-1}(\cdot)$ is the cumulative probability density function (cdf) of the standard normal distribution and
$$
\widehat{\sigma_t}^2=\Var(D_t^L-\widehat{D_t^L})
$$
is the variance of the forecast error for the lead time demand. In some articles (for example Chen et al., \cite{ch:dr:ry:si:00a}) $\widehat{\sigma_t}^2$ is defined more practically. That is, instead of the variance, one takes the sample variance of  $D_t^L-\widehat{D_t^L}$. This complicates the theoretical calculations somewhat and the estimation of $\widehat{\sigma_t}^2$ increases the bullwhip effect which can be deduced from the fact that Chen et al.'s \cite{ch:dr:ry:si:00a} formula is a lower bound for the bullwhip measure whereas we get an equality.

Note however, in our setting, even when demand is normally distributed, neither the inventory levels, nor the orders, are normally distributed. Rather the stochastic lead times create a multi-modal inventory distribution (as it did in Disney et al., \cite{di:ma:wa:wa:16}) and the lead time forecasting mechanism creates a multi-modal order distribution (which was not present in the setting considered by Disney et al., \cite{di:ma:wa:wa:16} as the lead time distribution was assumed to be known beforehand).  Thus in our case here, the $\mbox{\it TNS}$ must be set with
$$
\mbox{\it TNS}=F^{-1}\left(\frac{b}{b+h}\right)
$$
where $F^{-1}(\cdot)$ is the cdf of the inventory levels (arbitrary distribution).

Thus the order quantity $q_t$ placed
at the beginning of a period $t$ by the OUT policy is
\begin{equation}\label{qt}
q_t=S_t-S_{t-1}+D_{t-1}\,.
\end{equation}
Note that by   (\ref{eltd}),  (\ref{st}) and (\ref{qt}) the quantity of the order placed by the manufacturer to the supplier depends upon the supplier's lead time.

Our main purpose is to find $\Var q_t$ and then to calculate the following bullwhip ratio
$$
BM=\frac{\Var q_t}{\Var D_t}.
$$
This is one of the typical supply chain performance measurements (see e.g. Towill et al., \cite{to:zh:di:07}).
\begin{proposition}\label{varf}
The variance of the forecast error over the lead time demand does not depend on $t$ that is $\widehat{\sigma_t}^2=\widehat{\sigma}^2$.
\end{proposition}
\proof The variance of the forecast error is the expected value of a function of
$D_{t-n},$$\ldots,$ $D_{t-1},$ $D_{t},$ $D_{t+1},$$\ldots$ and $L_{t-m-L^+},$ $L_{t-m+1-L^+},$$\ldots,$ $L_{t-1-L^+}, L_t$ whose distribution is independent of $t$. The stationarity of the sequences $\{D_t\}_{t=-\infty}^\infty$ and $\{L_t\}_{t=-\infty}^\infty$ and their mutual independence yield the assertion.

Since the variance of the forecast error for the lead time demand is independent of $t$ we get from  (\ref{st}) and (\ref{qt}) that
\begin{equation}\label{qtf}
q_t=\widehat{D_t^L}-\widehat{D_{t-1}^L}+D_{t-1}\,.
\end{equation}
allowing us to derive the exact bullwhip expression.
\begin{theorem}\label{bmmt}
The measure of the bullwhip effect has the following form
\begin{equation}\label{BM}
\begin{split}
BM & =\frac{\Var q_t}{\Var D_t}\\ & =\frac{2\sigma_L^2}{n^2m^2}\left(m(1-\rho^n)+\frac{n(1+\rho)}{1-\rho}-\frac{(1+\rho^2)(1-\rho^n)}{(1-\rho)^2}\right)+\frac{2\sigma^2_L\mu^2_D}{\sigma^2_D m^2}+\left(\frac{2\mu_L^2}{n^2}+\frac{2\mu_L}{n}\right)(1-\rho^n)+1
\end{split}
\end{equation}
\end{theorem}
\proof The proof of Theorem 1 is given in Appendix 1.

\noindent
\emph{Remarks on Theorem 1}
The first summand \eqref{BM} describes the impact of
lead time variability, demand and lead time forecasting and the demand correlation. The second summand shows an impact of lead time forecasting, demand mean and variance and lead time variance on the bullwhip effect. The first two summands are not present in the constant lead time case. The third  term gives the amplification of the variance by demand forecasting, the demand correlation and the mean lead time.

If lead times are deterministic that is $L_t=L=const.$ then the bullwhip effect is described by
$$
BM_{L=const.}=
\left(\frac{2L^2}{n^2}+\frac{2L}{n}\right)(1-\rho^n)+1
$$
which coincides with Eq. 5 in Chen et al., \cite{ch:dr:ry:si:00a}. Note that Duc et al., \cite{du:lu:ki:08} also obtained the result of Chen et al., \cite{ch:dr:ry:si:00a} in a special case and
as an exact value (not a lower bound). Chen et al., \cite{ch:dr:ry:si:00a} obtain this expression as a lower bound because they define the error $\widehat{\sigma_t}$ (see (\ref{st})) as the sample variance of  $D_t^L-\widehat{D_t^L}$, indicating that the estimation of the variance of $D_t^L-\widehat{D_t^L}$ amplifies the bullwhip effect.

The following limits exist:
\begin{equation} \label{BMlimits1}
\lim_{n \to \infty} BM =1+ \frac{2 \mu_D^2 \sigma_L^2}{m^2 \sigma_D^2},
\end{equation}
\begin{equation} \label{BMlimits2}
\lim_{m \to \infty} BM =1+(1-\rho^n)\left(\frac{2\mu_L^2}{n^2}+\frac{2\mu_L}{n}\right),
\end{equation}
\begin{equation} \label{BMlimits3}
 \lim_{\{n,m\} \to \infty} BM =1.
\end{equation}

It is easy to see from (\ref{BM}) that bullwhip is strictly decreasing in $m$, but this is not true for $n$ as there is an odd-even effect in $n$ for negative $\rho$. When $n=1$ then the BM is a linear function in $\rho$ as
\begin{equation} \label{BMn1}
BM_{n \to 1}=\frac{\rho 2 \sigma _D^4 \left(\sigma _L^2-m \left(m \mu _L \left(\mu _L+1\right)+\sigma _L^2\right)\right)}{m^2}+\frac{2 \mu _D^2 \sigma _D^2 \sigma _L^2+m \sigma _D^4 \left(2 m \mu _L \left(\mu _L+1\right)+2 \sigma _L^2+m\right)}{m^2},
\end{equation}
which always has a negative gradient in $\rho$ (unless $\mu_L=0$ and $m=1$, in which case the gradient is zero).

For i.i.d. demand the following bullwhip measure exists
\begin{equation} \label{BMiid}
\begin{split}
BM_{iid} & =1+\frac{2 \mu _L^2}{n^2}+\frac{2 \mu _L}{n}+\frac{2 \mu _D^2 \sigma _L^2}{m^2 \sigma _D^2}-\frac{2 \sigma _L^2}{m^2 n^2}+\frac{2 \sigma _L^2}{m^2 n}+\frac{2 \sigma _L^2}{m n^2}\\
& =1+\frac{2 \mu _D^2 \sigma _L^2}{m^2 \sigma _D^2}+\frac{2 \sigma _L^2 (m+n-1)}{m^2 n^2}+\frac{2 \mu _L \left(\mu _L+n\right)}{n^2}
\end{split}
\end{equation}
which is strictly decreasing in $n$ and $m$ and the result is consistent with Michna and Nielsen \cite{mi:ni:13}. The derivative of the bullwhip measure in (\ref{BM}) at $\rho =0$ is
\begin{equation}\label{DBM0}
\frac{dBM }{d\rho}\bigg|_{\rho = 0}=\frac{4 (n-1) \sigma _L^2}{m^2 n^2}
\end{equation}
which is always positive when $n>1$.

As $\rho\rightarrow 1$ then the following expression defines the bullwhip measure
\begin{equation} \label{BM1}
BM_{\rho\to 1}=1+\frac{2 \sigma _L^2 \left(\mu _D^2+\sigma _D^2\right)}{m^2 \sigma _D^2}
\end{equation}
which is independent of $n$ and decreasing in $m$.
Notice $BM_{iid}\geq BM_{\rho \to 1}$ if
\begin{equation} \label{BMcond}
n\leq\frac{\sigma_L^2+m^2\mu_L +\sqrt{(\sigma_L^2+m^2\mu_L)^2+4\sigma_L^2(\sigma_L^2(m-1)+m^2\mu_L^2)}}{2\sigma_L^2}
\end{equation}
holds.
Eq. (\ref{BMcond}) together with (\ref{DBM0}) provides a sufficient (but not necessary) condition for the presence of at least one stationary point in the region $0<\rho<1$ if $n\geq 2$ (see Fig. \ref{n5m2}). Notice that if the lead time is a constant then $BM_{iid}>BM_{\rho \to 1}$.

If $\rho\rightarrow -1$ then
\begin{equation} \label{BMminus1}
BM_{\rho \to -1}=1+\frac{2 \mu _D^2 \sigma _L^2}{m^2 \sigma _D^2}-\frac{(2 m-1) \left((-1)^n-1\right) \sigma _L^2}{m^2 n^2}-\frac{2 \left((-1)^n-1\right) \mu _L \left(\mu _L+n\right)}{n^2}
\end{equation}
which is decreasing in $m$, but the odd-even impact of $n$ can be clearly seen.  When $n$ is even then
\begin{equation} \label{DBMminus1}
BM_{\rho \to -1, \text{ even }n}=1+\frac{2 \mu _D^2 \sigma _L^2}{m^2 \sigma _D^2}\,.
\end{equation}
Numerical investigations (see Figs. \ref{n6m2}, \ref{n16m2}, \ref{n6m20}, and \ref{n22m20}) seem to suggest that there are no stationary points in the region $-1<\rho<0$ when $n$ is even, but we remain unable to prove so. However this is congruent with our previous results that $BM_{iid}>BM_{\rho \to -1, \text{ even }n}$ and $\frac{dBM}{d\rho}\big|_{\rho = 0}>0$.

When $n$ is odd then
\begin{equation} \label{BMminus1odd}
BM_{\rho \to -1, \text{ odd }n}=1+\frac{2 \mu _D^2 \sigma _L^2}{m^2 \sigma _D^2}+\frac{2 (2 m-1) \sigma _L^2}{m^2 n^2}+\frac{4 \mu _L \left(\mu _L+n\right)}{n^2}\,.
\end{equation}

Finally, $BM_{iid}\leq BM_{\rho \to -1, \text{ odd }n}$ if
\begin{equation} \label{minus1cond}
m\geq \frac{\sigma_L\sqrt{\sigma_L^2+4n\mu_L(\mu_L+n)}-\sigma_L^2}{2\mu_L(\mu_L+n)}\,.
\end{equation}
When (\ref{minus1cond}) holds and $n>1$ there must be at least one stationary point between $-1<\rho<0$ because of the positive derivative at $\rho=0$, see (\ref{DBM0}). Note this is again a sufficient, but not a necessary condition. Extensive numerical investigations (see Figs. \ref{n5m2} and \ref{n15m2}) suggest that only one stationary point exists in this area though we can not prove it. Moreover
for large $m$ the derivative at $\rho=0$ is almost zero, see (\ref{DBM0}) and Figs. \ref{n5m20} to \ref{n22m20}.

\section{Numerical investigations}

Let us further investigate the influence of the demand correlation on the bullwhip effect by analyzing some concrete numerical examples. We plot the bullwhip effect measure as a function of the demand correlation parameter $\rho$. Fixing
$\mu_D=20$, $\sigma_D=4$, $\mu_L=10$, $\sigma_L=5$ we depict the bullwhip measure in four different scenarios. That is when: $n$ and $m$ are small, one of them
is small and the other is large and both are large. If $n$ is small, we need to distinguish
two further cases; that is, whether $n$ is even or odd.

\begin{figure}[!h]
\centering
\begin{minipage}{.51\textwidth}
  \centering
  \includegraphics[width=1\linewidth]{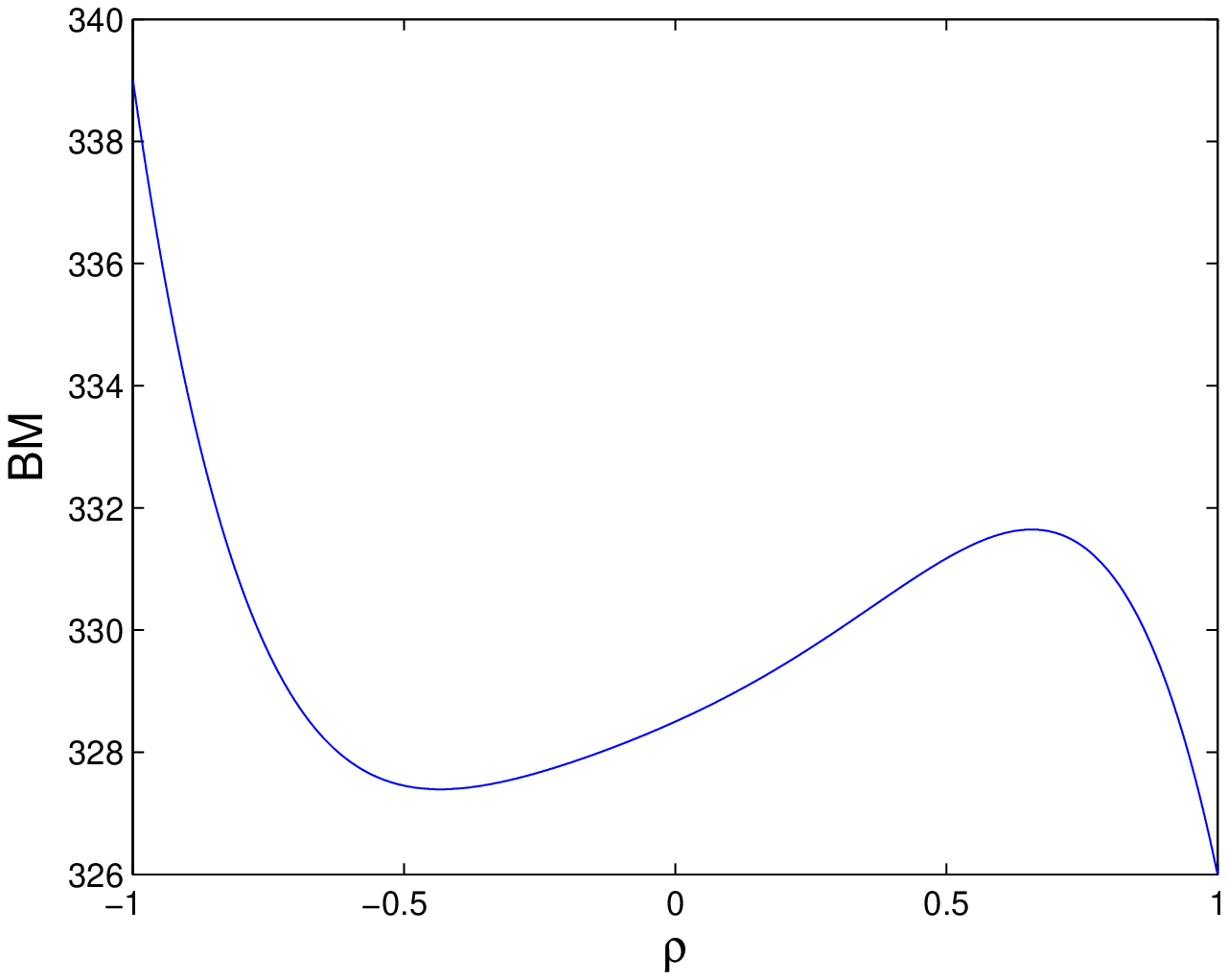}
  \caption{Bullwhip when $n=5$ and $m=2$}
  \label{n5m2}
\end{minipage}%
\begin{minipage}{.51\textwidth}
  \centering
  \includegraphics[width=1\linewidth]{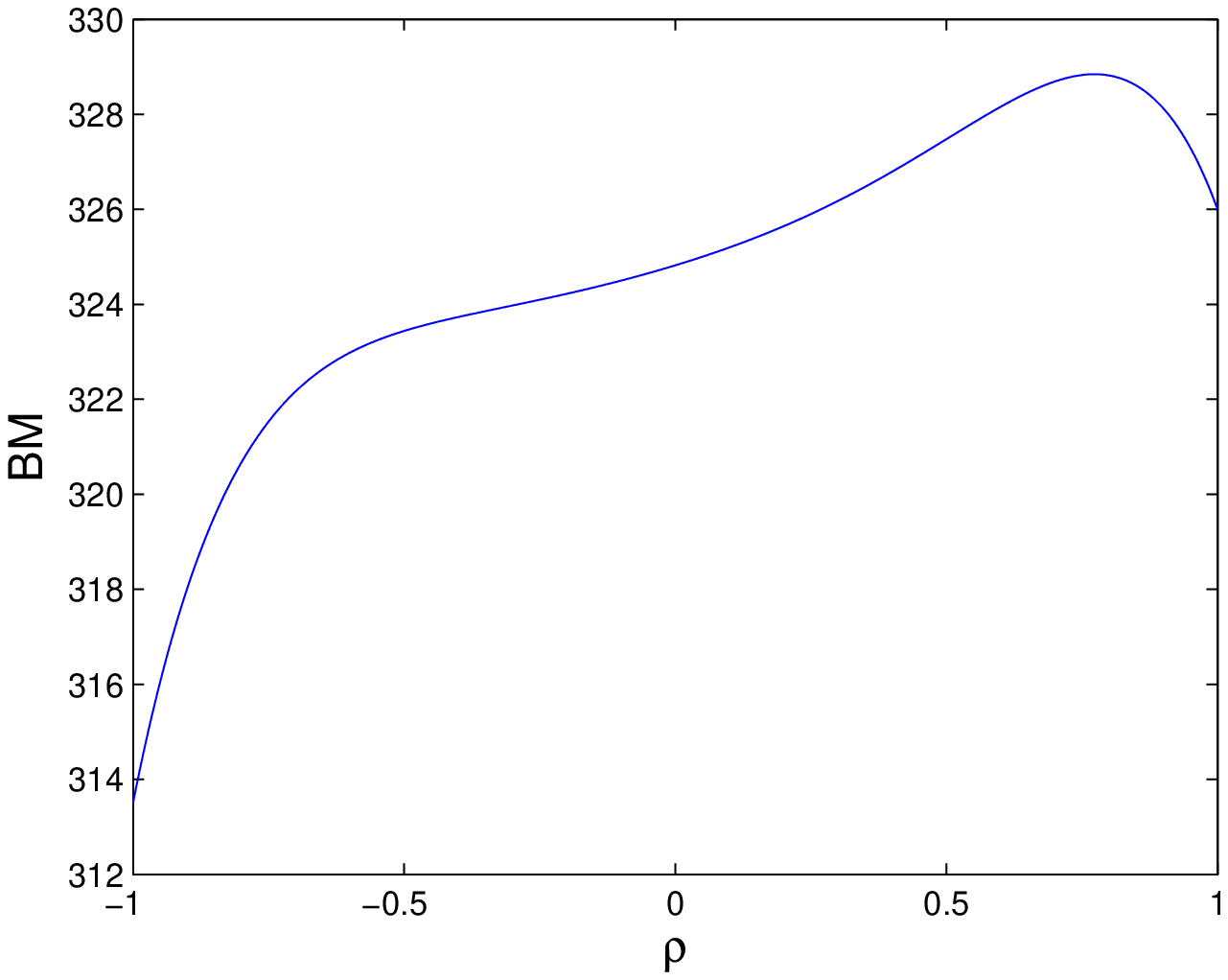}
  \caption{Bullwhip when $n=6$ and $m=2$}
  \label{n6m2}
\end{minipage}
\end{figure}

Thus if $n=5$ and $m=2$ (small and odd $n$) the bullwhip measure has a minimum at $\rho\approx -0.5$ and $\rho= 1$ and a maximum at $\rho= -1$ and
$\rho\approx 0.7$ (see Fig. \ref{n5m2}). The bullwhip measure behavior for $\rho= -1$ and $\rho= 1$ can be predicted by taking the limit as the AR(1) model is well-defined when $-1<\rho<1$. In Duc et al., \cite{du:lu:ki:08} the minimal value of $BM$ is attained for $\rho$ near $-0.6$ or $-0.7$ and the maximal value of $BM$ is for $\rho$ around
$0.6$ or $1$. Their results are very close to ours if $n$ and $m$ are small and $n$ is odd (see Fig. \ref{n5m2}) because their model does not predict the lead time, corresponding to the small values of $n$ and $m$ in our model.

For $n=6$ and $m=2$ (small and even $n$) we observe a different behavior. Specifically, the smallest value of the bullwhip effect is attained for $\rho= -1$ and the largest for $\rho\approx 0.75$ (see Fig. \ref{n6m2}). This concurs with Kahn \cite{ka:87} who revealed positively correlated demands result in the bullwhip effect. We also notice that the bullwhip effect is very large if $n$ and $m$ are small.

\begin{figure}[!h]
\centering
\begin{minipage}{.51\textwidth}
  \centering
  \includegraphics[width=1\linewidth]{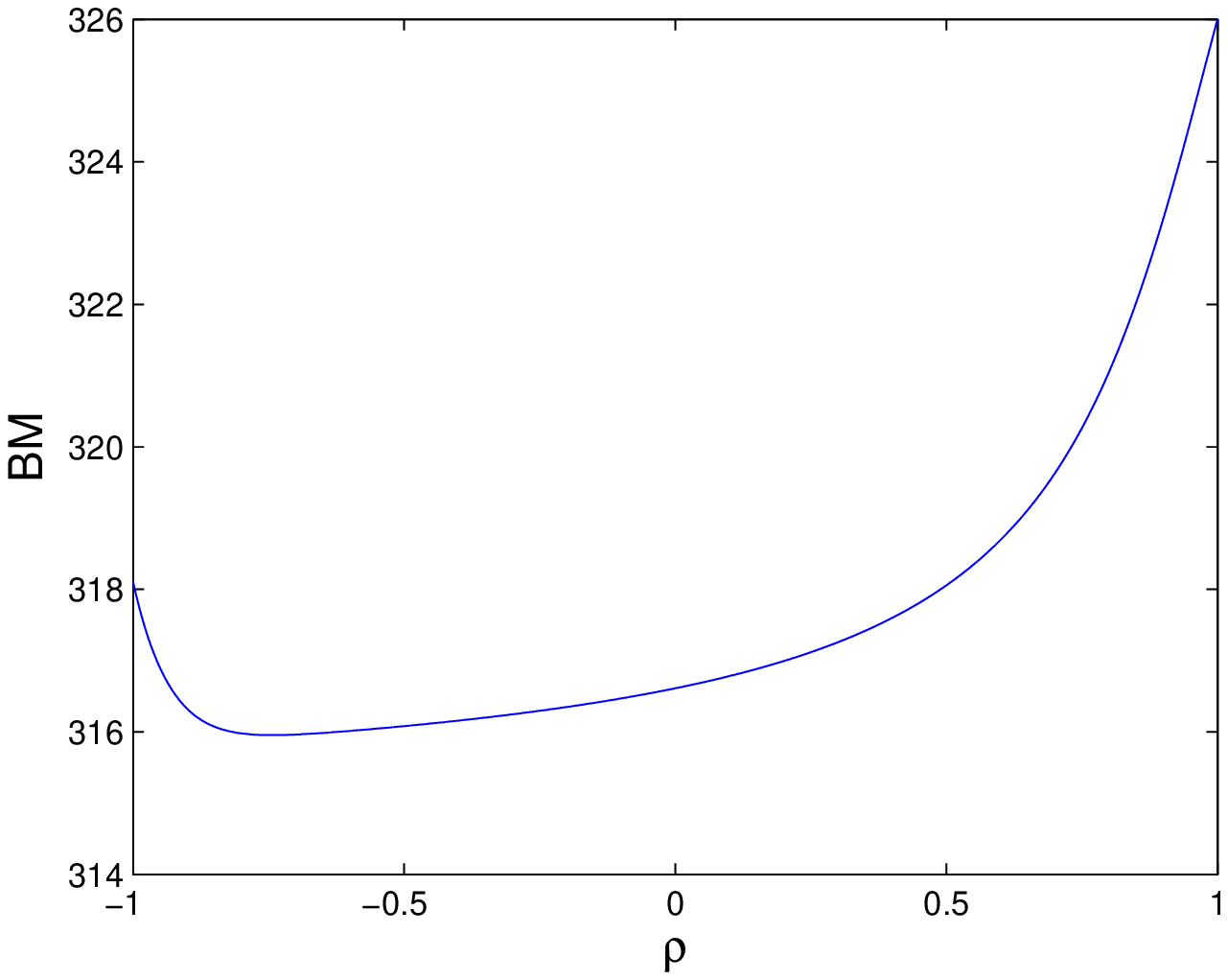}
  \caption{Bullwhip when $n=15$ and $m=2$}
  \label{n15m2}
\end{minipage}%
\begin{minipage}{.51\textwidth}
  \centering
  \includegraphics[width=1\linewidth]{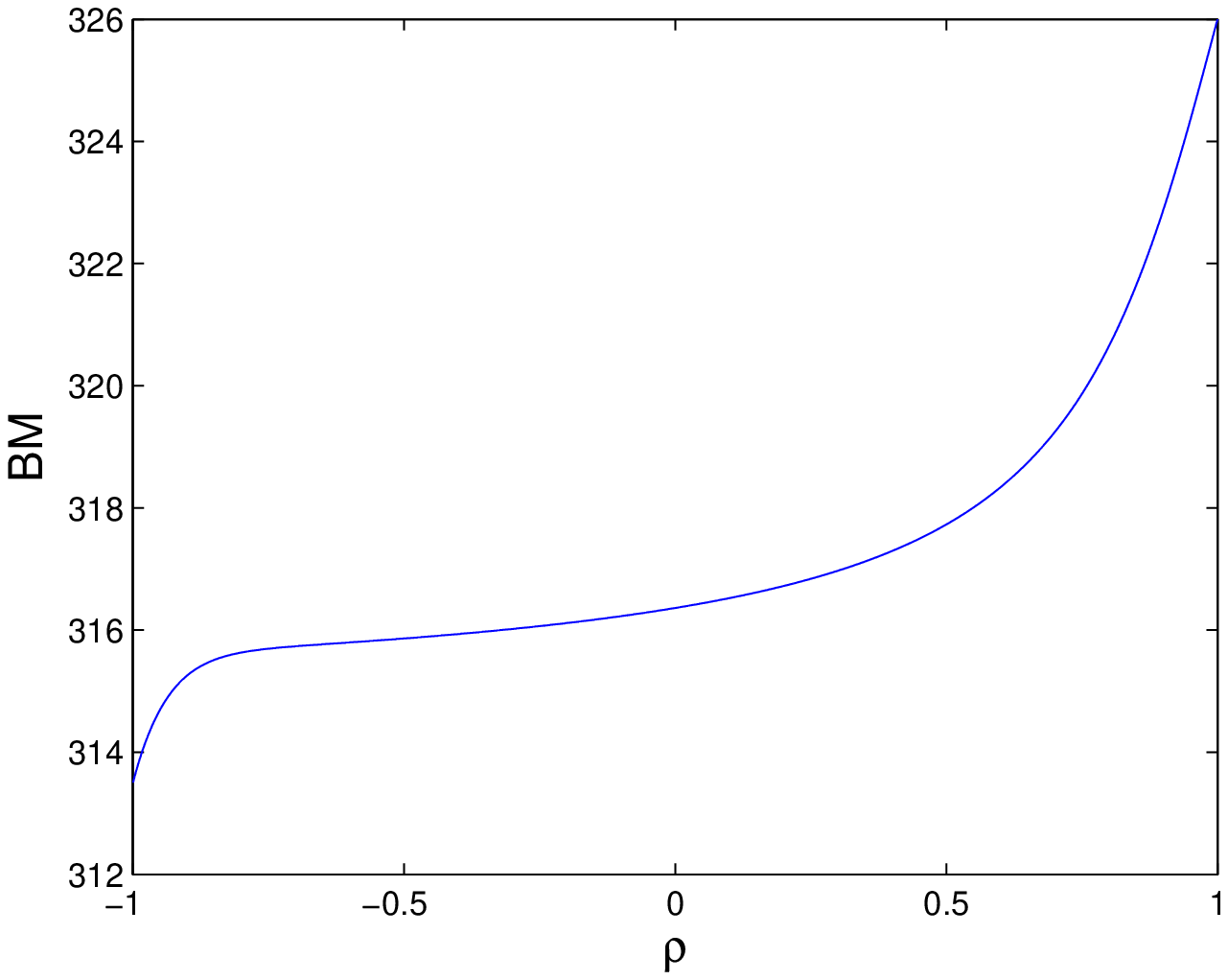}
  \caption{Bullwhip when $n=16$ and $m=2$}
  \label{n16m2}
\end{minipage}
\end{figure}

The situation changes if $n$ is large and $m$ is  small (see Figs. \ref{n15m2} and \ref{n16m2}). Then the bullwhip measure is almost an increasing function of the demand correlation
except for odd $n$ and $\rho$ close to $-1$ where we observe a minimum. Moreover bullwhip
increases quite slowly in the region of $-0.8<\rho<0.5$. The odd-even
effects in $n$ are now much less noticeable. These observations are consistent with our  theoretical analysis in the previous section.

\begin{figure}[!h]
\centering
\begin{minipage}{.51\textwidth}
  \centering
  \includegraphics[width=1\linewidth]{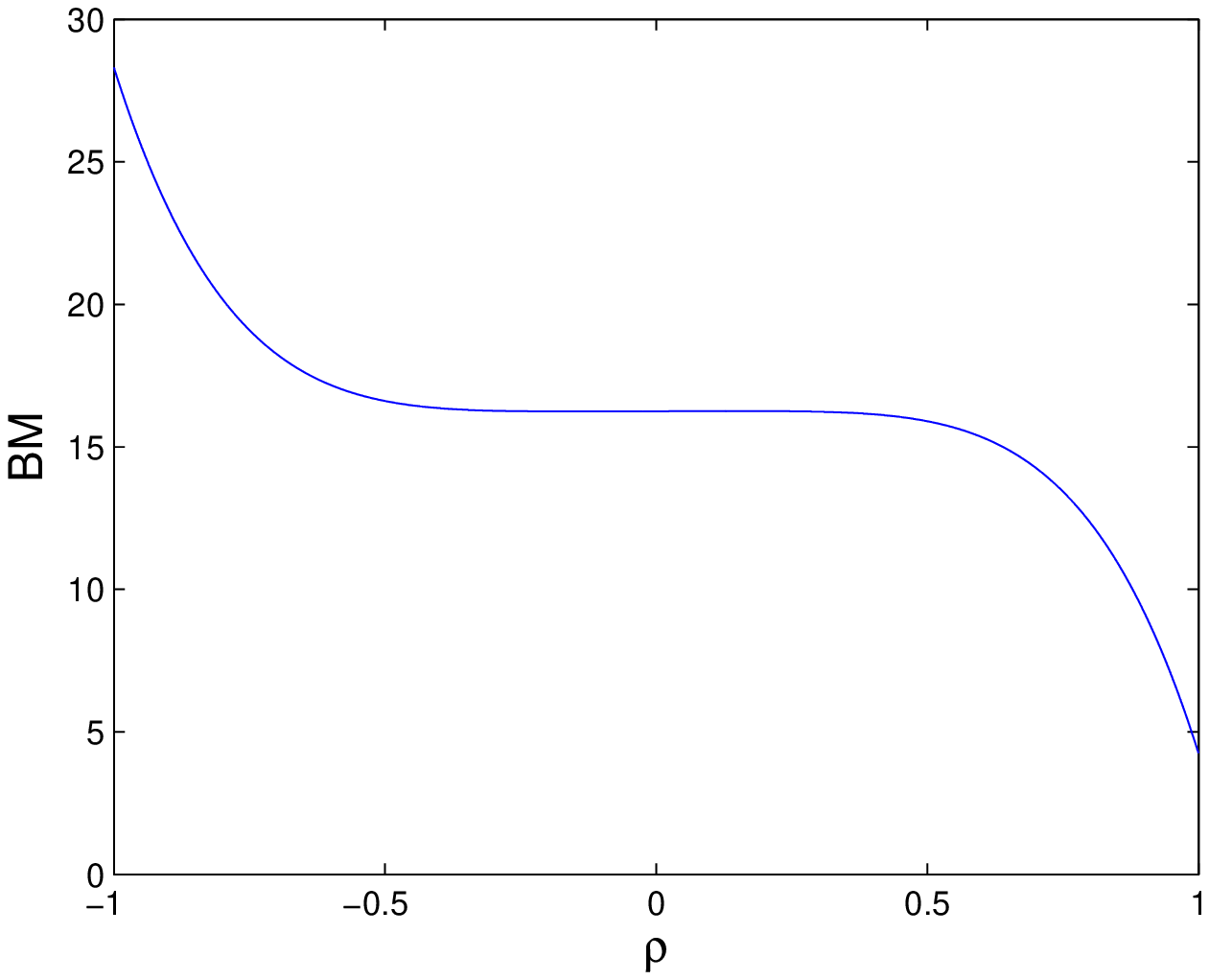}
  \caption{Bullwhip when $n=5$ and $m=20$}
  \label{n5m20}
\end{minipage}%
\begin{minipage}{.51\textwidth}
  \centering
  \includegraphics[width=1\linewidth]{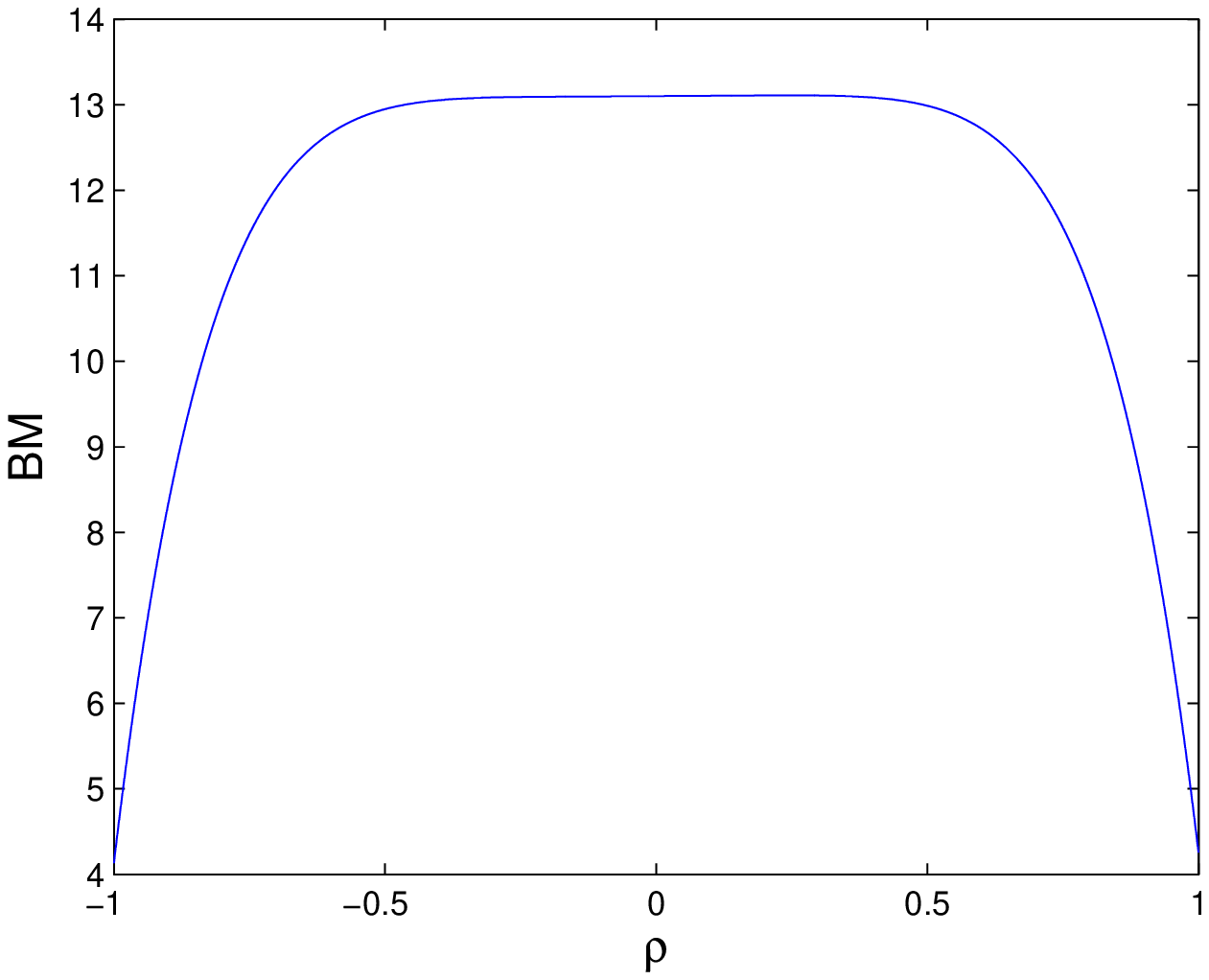}
  \caption{Bullwhip when $n=6$ and $m=20$}
  \label{n6m20}
\end{minipage}
\end{figure}

\begin{figure}[!h]
\centering
\begin{minipage}{.51\textwidth}
  \centering
  \includegraphics[width=1\linewidth]{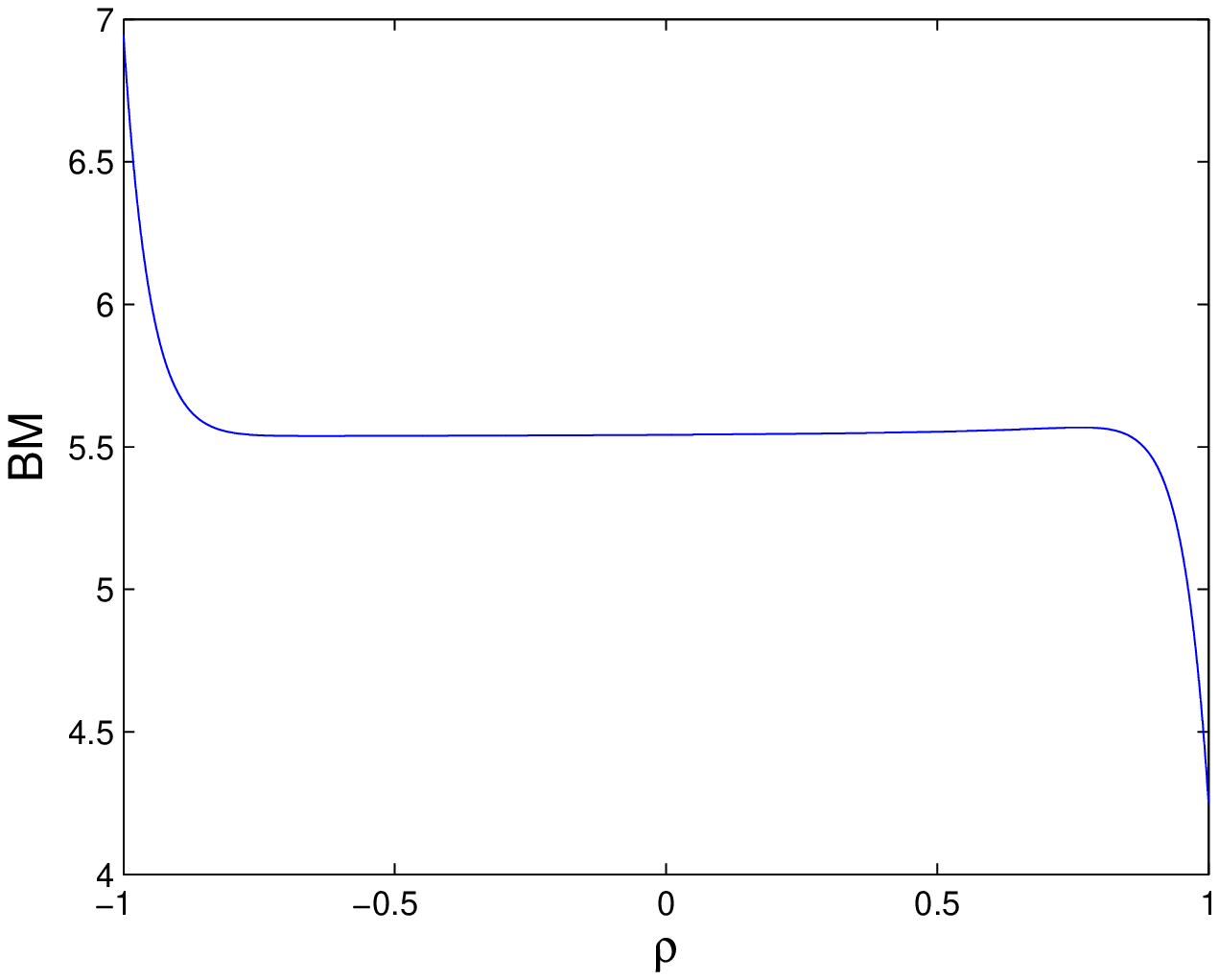}
  \caption{Bullwhip when $n=21$ and $m=20$}
  \label{n21m20}
\end{minipage}%
\begin{minipage}{.51\textwidth}
  \centering
  \includegraphics[width=1\linewidth]{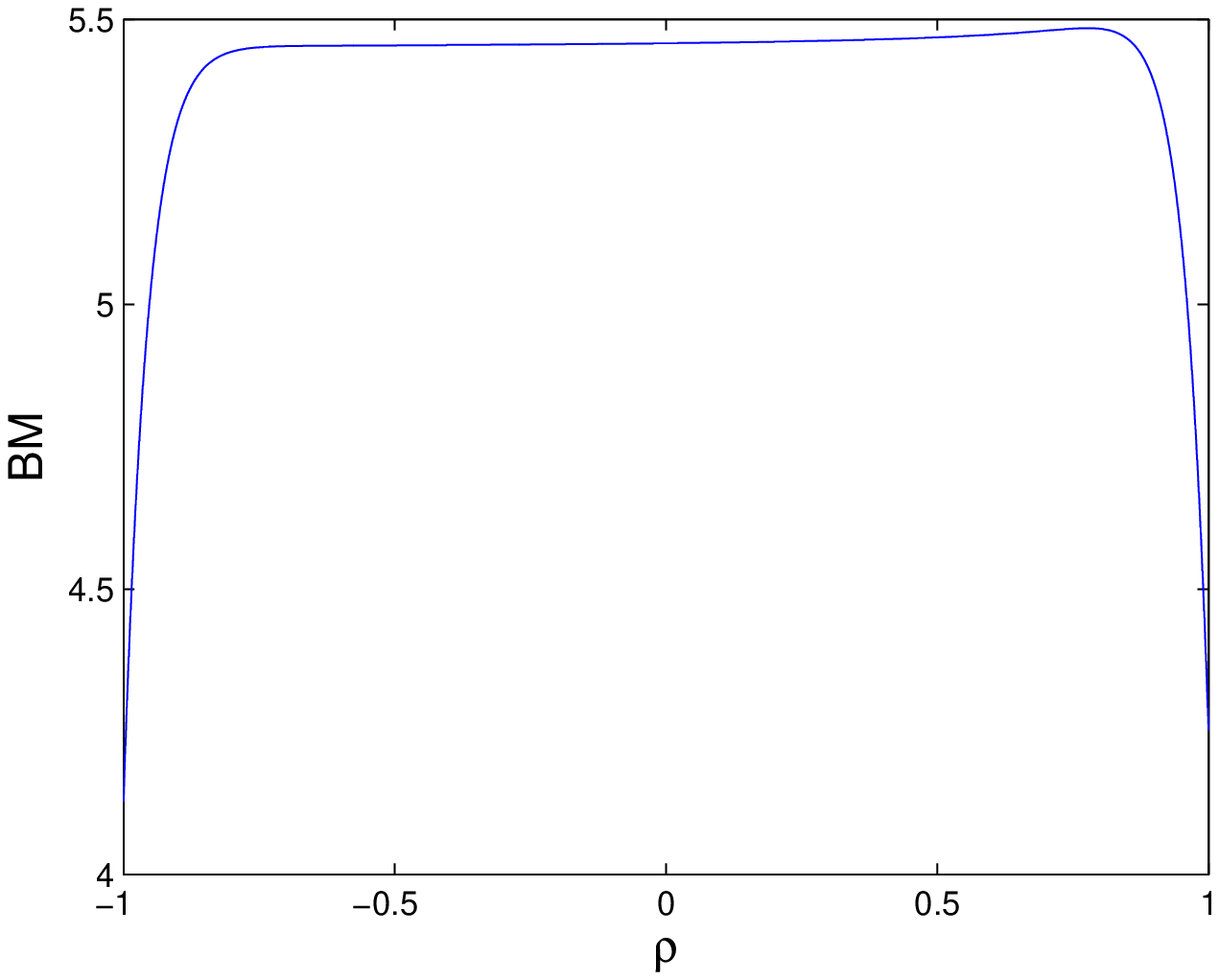}
  \caption{Bullwhip when $n$=\,22 and $m=20$}
  \label{n22m20}
\end{minipage}
\end{figure}

As $m$, the number of periods used in the moving average forecast of the lead time increases, the bullwhip effect becomes independent of the demand correlation $\rho$ regardless
of the number of periods used in the moving average forecast of demand, $n$. That is, bullwhip remains almost constant except near $\rho=\{-1,1\}$. Figs. \ref{n5m20} to \ref{n22m20} confirm this independence for the cases when $n=5$, $n=6$, $n=21$, $n=22$ and $m=20$. This is caused by the first summand of (\ref{BM}) which vanishes as $m\rightarrow\infty$ or $n\rightarrow\infty$ and the third summand is rather insensitive to $\rho$.
For $\rho$ close to $-1$ or $1$ the bullwhip effect can dramatically increase or decrease. Moreover much less bullwhip is generated with large values of $n$ and $m$.  Reduced bullwhip with large $m$ is congruent with the results of Disney et al., \cite{di:ma:wa:wa:16} and Wang and Disney \cite{wa:di:15}.


\section {Conclusions and further research opportunities}
We quantified the bullwhip effect when demands and lead times must be forecasted. Demand and lead time forecasting are necessary when placing an order if demands and lead times are stochastic. We have confirmed, extended and sharpened the conclusion of Michna and Nielsen \cite{mi:ni:13} that lead time forecasting is a major cause of the bullwhip effect. We assumed that demands constitute a first order autoregressive process and we obtained quantitative results which link bullwhip and the demand correlation when demands and lead times are to be predicted separately. We conclude that how one goes about forecasting demand and lead time is important as it can cause significant amounts of bullwhip. Moreover the dependence of the bullwhip measure on the demand correlation parameter is different according to the forecasting parameters used to make lead time and demand predictions.  In particular, an even number of data points in the moving average demand forecast can significantly reduce bullwhip when demand has a strong negative correlation.

Future bullwhip research could be focused on the impact of lead time forecasting under the assumption that lead times are correlated, either temporally, or with other system states such as customer demand.  For example if large orders lead to long lead times, there is a correlation between the lead time and the order size and this dependence should be captured somehow.  This seems to be difficult to quantify analytically.

Other opportunities lie in studying the impact of different forecasting methods for lead times and demands (see Zhang \cite{zh:04} for the case of demand forecasting). Another big challenge is to quantify bullwhip in the presence of unrealized previous orders when placing an order. More precisely, forecasting the most recent lead times when some orders are not yet received will distort the lead time distribution and have an impact on bullwhip.
The quantification of this issue seems to be a difficult task, but it will become important when the lead times are temporally correlated.

An important challenge is the investigation of the variance amplification of orders and the variance amplification of inventory levels simultaneously because an improper focus on bullwhip reduction can amplify the variability of inventory levels (see Chen and Disney \cite{ch:di:07}, Devika et al., \cite{de:ja:ha:kh:14}, Disney at el. \cite{di:ma:wa:wa:16} and Wang and Disney \cite{wa:di:15}) which can be as harmful as the bullwhip. Moreover, the POUT replenishment policy, Disney and Towill \cite{di:to:03}, should be investigated under the assumption of lead time forecasting.

\section{Appendix}
{\bf Proof of Th. \ref{bmmt}.}
We apply the law of total variance to find variance of $q_t$. Namely, let us put
$$
L=(L_{t-1-L^+}, L_{t-2-L^+},\ldots, L_{t-1-m-L^+})
$$
and then
\begin{equation}\label{varl}
\Var q_t=\Var(\Exp(q_t|L))+\Exp(\Var(q_t|L))\,.
\end{equation}
Using (\ref{qtf}) it can be seen that
\begin{eqnarray}
q_t&=&\widehat{D_t^L}-\widehat{D_{t-1}^L}+D_{t-1}\nonumber\\
&=&\widehat{L_t}\widehat{D_t}-\widehat{L_{t-1}}\widehat{D_{t-1}}+D_{t-1}\label{qtfe}
\end{eqnarray}
revealing that $\Exp q_t=\Exp D_t=\mu_D$. Moreover from the second expression for $q_t$ it follows that
$$
\Exp(q_t|L)=(L_{t-1-L^+}-L_{t-1-m-L^+})\frac{\mu_D}{m}+\mu_D
$$
which gives
\begin{equation}\label{ftqt}
\Var(\Exp(q_t|L))=\frac{2\sigma^2_L\mu^2_D} {m^2}\,.
\end{equation}
To calculate the conditional variance of $q_t$ we express it as a function of $D_{t-1-n}$ and the error terms $\epsilon_{t-n}, \epsilon_{t-n+1}, \ldots, \epsilon_{t-1}$ which are mutually independent. Thus by (\ref{qtfe}) and (\ref{ard}) we get
\begin{eqnarray*}
q_t
&=&\left(\frac{\widehat{L_t}}{n}+1\right)D_{t-1}+\frac{L_{t-1-L^+}-L_{t-1-m-L^+}}{nm}\sum_{k=2}^n D_{k-2}
-\frac{\widehat{L_{t-1}}}{n}D_{t-1-n}\\
&=&\left(\frac{\widehat{L_t}}{n}+1\right)\mu_D(1-\rho^n)+\frac{(L_{t-1-L^+}-L_{t-1-m-L^+})\mu_D}{nm}\left(n-1-\frac{\rho(1-\rho^{n-1})}{1-\rho}\right)\\
&&+\left[\left(\frac{\widehat{L_t}}{n}+1\right)\rho^n+\frac{(L_{t-1-L^+}-L_{t-1-m-L^+})\rho(1-\rho^{n-1})}{nm(1-\rho)}-\frac{\widehat{L_{t-1}}}{n}\right]D_{t-1-n}\\
&&+\sum_{k=1}^n\left[\left(\frac{\widehat{L_t}}{n}+1\right)\rho^{k-1}+\frac{(L_{t-1-L^+}-L_{t-1-m-L^+})(1-\rho^{k-1})}{nm(1-\rho)}\right]\epsilon_{t-k}
\end{eqnarray*}
which gives
\begin{equation}\label{cvarqt}
\Var(q_t|L)=\sigma^2_D C^2_1+\sigma^2\sum_{k=1}^n C^2_{2, k}\,,
\end{equation}
where
$$
C_1=\left(\frac{\widehat{L_t}}{n}+1\right)\rho^n+\frac{(L_{t-1-L^+}-L_{t-1-m-L^+})\rho(1-\rho^{n-1})}{nm(1-\rho)}-\frac{\widehat{L_{t-1}}}{n}
$$
and
$$
C_{2, k}=\left(\frac{\widehat{L_t}}{n}+1\right)\rho^{k-1}+\frac{(L_{t-1-L^+}-L_{t-1-m-L^+})(1-\rho^{k-1})}{nm(1-\rho)}\,.
$$
Thus we get
$$
\Exp C_1=\left(\frac{\mu_L}{n}+1\right)\rho^n-\frac{\mu_L}{n}
$$
and
$$
\Exp C_{2, k}=\left(\frac{\mu_L}{n}+1\right)\rho^{k-1}\,.
$$
To calculate $\Exp\Var(q_t|L)$ it is necessary to find $\Exp C^2_1$ and $\Exp C^2_{2, k}$ by (\ref{cvarqt}). We compute them finding variance and adding the square of the first moment. Thus, to obtain the variance of $C_1$ and $C_{2, k}$, we express them as a sum of independent random variables that is
$$
C_1=\frac{\rho^n-1}{nm}\sum_{k=2}^m{L_{t-k-L^+}}+\frac{\rho(1-\rho^n)}{(1-\rho)nm}L_{t-1-L^+}-\frac{1-\rho^n}{(1-\rho)nm}L_{t-1-m-L^+}+\rho^n
$$
and
$$
C_{2, k}=\frac{\rho^{k-1}}{nm}\sum_{k=2}^m{L_{t-k-L^+}}+\frac{1-\rho^k}{(1-\rho)nm}L_{t-1-L^+}-\frac{1-\rho^{k-1}}{(1-\rho)nm}L_{t-1-m-L^+}+\rho^{k-1}\,.
$$
Hence we obtain
$$
\Var C_1=\frac{(1-\rho^n)^2\sigma^2_L}{n^2m^2}\left(m+\frac{2\rho}{(1-\rho)^2}\right)
$$
and
$$
\Var C_{2, k}=\frac{\sigma^2_L}{n^2m^2}\left[\rho^{2(k-1)}(m-1)+\left(\frac{1-\rho^k}{1-\rho}\right)^2+\left(\frac{1-\rho^{k-1}}{1-\rho}\right)^2\right]\,.
$$
So we get
\begin{equation}\label{ec21}
\Exp C^2_1=\frac{(1-\rho^n)^2\sigma^2_L}{n^2m^2}\left(m+\frac{2\rho}{(1-\rho)^2}\right)
+\left[\left(\frac{\mu_L}{n}+1\right)\rho^n-\frac{\mu_L}{n}\right]^2
\end{equation}
and
\begin{eqnarray*}
\Exp C^2_{2, k}&=&\left(\frac{\sigma^2_L(m-1)}{n^2m^2}+\left(\frac{\mu_L}{n}+1\right)^2+
\frac{\sigma^2_L(\rho^2+1)}{n^2m^2(1-\rho)^2}\right)\rho^{2(k-1)}\\
&&-\frac{2\sigma^2_L(\rho+1)}{n^2m^2(1-\rho)^2}\rho^{k-1}+\frac{2\sigma^2_L}{n^2m^2(1-\rho)^2}\,.
\end{eqnarray*}
Summing the last expression we obtain
\begin{eqnarray}
\sum_{k=1}^n\Exp C^2_{2, k}&=&\left(\frac{\sigma^2_L(m-1)}{n^2m^2}+\left(\frac{\mu_L}{n}+1\right)^2+
\frac{\sigma^2_L(\rho^2+1)}{n^2m^2(1-\rho)^2}\right)\frac{1-\rho^{2n}}{1-\rho^2}\nonumber\\
&&-\frac{2\sigma^2_L(\rho+1)}{n^2m^2(1-\rho)^2}\frac{1-\rho^n}{1-\rho}+\frac{2\sigma^2_L}{nm^2(1-\rho)^2}\,.\label{sumc2k}
\end{eqnarray}
Plugging (\ref{sumc2k}), (\ref{ec21}), (\ref{cvarqt}), (\ref{ftqt}) into (\ref{varl}) yields the formula from the assertion after a simple algebra.

\subsection*{Acknowledgments}
The first author acknowledges support by the National Science Centre grant 2012/07/B//HS4/00702.

\end{document}